\documentclass[conference]{IEEEtran}

\usepackage[utf8]{inputenc}
\usepackage{graphicx}
\usepackage{amsthm,amsmath}
\usepackage{textcomp}
\usepackage{xcolor}
\usepackage{verbatim}
\usepackage{amsmath}
\usepackage{amssymb}
\usepackage{bm}
\usepackage{bbm}
\usepackage{makecell}
\usepackage[noadjust]{cite}

\usepackage[subrefformat=parens,labelformat=parens,caption=false,font=footnotesize]{subfig}

\ifCLASSINFOpdf
\else
\fi
\begin{document}
%

\title{Toward 6G: From New Hardware Design to Wireless Semantic and Goal-Oriented Communication Paradigms}

\author{\IEEEauthorblockN{Emilio Calvanese Strinati, Didier Belot, Alexis Falempin, Jean-Baptiste Dor\'{e} \\ CEA-Leti, Grenoble, France }}
\maketitle

\begin{abstract}
Several speculative visions are conjecturing on what  6G services will be able to offer at the horizon of 2030. Nevertheless, the 6G design process is at its preliminary stages. The reality today is that hardware, technologies and new materials required to effectively meet the unprecedented performance targets  required for future 6G services and network operation, have not been designed, tested or even do not exist yet. Today, a solid vision on the cost-benefit trade-offs of machine learning and artificial intelligence support for 6G network and services operation optimization is missing. This includes the possible support from hardware efficiency, operation effectiveness and, the immeasurable cost due to data acquisition-transfer-processing. 
The contribution of this paper is three-fold. This is the first paper deriving crucial 6G key performance indicators on hardware and technology design. Second, we present a new hardware technologies design methodology conceived to enable the effective software-hardware components integration required to meet the challenging performance envisioned for future 6G networks. Third, we suggest a paradigm shift towards \textit{goal-oriented} and \textit{semantic communications}, in which a totally  new opportunity of joint design of  hardware, artificial intelligence and effective communication is offered. The proposed vision is consolidated by our recent results on hardware, technology and machine learning performance. 


\end{abstract}


%

\IEEEpeerreviewmaketitle

\section{Introduction}
Academic and industrial research have aggressively started worldwide to drawing and discussing the potential offered by the next sixth generation (6G) of wireless communications systems. Many visions are under discussion today \cite{Calvanese6GVTM2019, Letaief20196GRoadMap, Yang2019UseCase6G, CalvaneseGOWSC2021, Calvanese2020Sky6G, CalvaneseRISE-6G2021,CalvaneseHexaX2021}. Those visions advocate for the need of designing and engineering a totally new generation of connect-compute-control systems that cannot be efficiently enabled by the current development of 5G networks. The ambition is to enable the immersive and native convergence of physical and cyber words. Distributed artificial intelligence (AI) will support effective network and services operation, but AI will also evolve and regenerate autonomously. 6G will indeed natively enable interactions between AI agents that will develop capabilities, transport information and, offer support for a new class of services: the \textit{semantic services} \cite{CalvaneseGOWSC2021}. Semantic services  will support many types of new applications and use cases not bounded to human to human (H2H) interactions but broadly will address the seamless connection and intertwining of different kinds of intelligence, both natural and artificial. Indeed with 6G it will be possible to effectively support, for the first time, knowledge sharing between the interacting parties:  H2H, human to machine (H2M) and machine to machine interactions (M2M). Such revolution will be possible thanks to perceived infinite-link-capacity, zero-latency, zero-energy, ultra reliable transfer and semantically enhanced mining of data that 6G will offer at both terrestrial and non-terrestrial networks \cite{Calvanese2020Sky6G}. This will require revolutionary energy efficient solutions at both system and technology levels, the exploitation of new materials and, the native adoption of artificial intelligence (AI).

Since the definition of the first communication generation (1G), the process required to define and standardize  \textit{the next generation} is repeated. This process requires between eight to ten years from first research investigations, visions, engineered validations. Since the second generation (2G) of communication systems, we have witnessed an endless antagonism between \textit{clean slate} technology inclusion and progressive \textit{evolution} of the current and past generations. This generation design cycle repeats: the \textit{next generation} is defined at the intersection between the availability of new technological breakthrough which will eventually be functioning when the prototyping and testing phase validate candidate solutions for the inclusion of technologies into the new standard and, the business oriented push for new service, use cases and applications that cannot be conveniently offered by the previous generation.

Continuous quest for business growth repetitively orients \textit{next generations} design to push technology, system design and architectures to  scale up performance such as requiring higher throughput, reduced latency, increased reliability and ubiquitous service coverage. The current vision on 6G follows the same dynamic: the new generation has to offer a factor of 100, 1000 or even more of improvement on envisioned key performance indicator (KPIs) metrics. 
Indeed, 6G continues the race started already in \textit{1G}, targeting to offer higher link and system capacity compared to previous generations. The accepted direction is to explore new spectrum horizons and target higher spectral efficiency for wireless communications. To this end, new challenging research axes explore the use of sub-Terahertz and visible light spectrum \cite{Calvanese6GVTM2019, dore2020SubTHz, Chowdhury2020, Rappaport2019}. Once, those frequencies were thought of as unusable frequencies. Today, we are investigating on how to effectively use frequencies beyond 90 GHz in a real communication system. This requires to face multi-fold challenges due to both the many propagation and reception issues, the new short-communication range paradigm and, the needed for new hardware and solid-state technology design. This includes (i) the severe path loss due to atmospheric absorption and possible loss of line-of-sight (LoS) that might cause even very severe blocking, causing momentary interruption of the communication link; (ii) advanced antennas arrays design to counteract the propagation losses thought high gain directional communications; (iii) the design of new wave-forms \cite{Sarieddeen_2020} 
to increase the effective use of such high frequency bands; (iv) new signal processing techniques, potentially empowered with AI mechanisms, new Radio Frequency (RF) channel optimization, connect-compute resource allocation, (v) the fundamental design and development of new hardware and solid-state technologies to bridge desired performance to reality, facing issues in terms of hardware compactness, isolation, selectability, linearity as well as, losses in transmission, reception and processing and, severe losses in RF-to-antennas connection. In our view, it will be required also to investigate revolutionary co-design and integration of radio frequency (RF) and antennas as well as innovative use of material hybridization techniques.
Concerns on sustainability and environmental impact of the new generation are shaping up new KPIs for 6G. As a consequence, 6G vision is also focused on the \textit{sustainability} of networks and services. The concept of energy efficiency has already been brought to the attention of communication systems at the conception of 4G long term evolution with the main concepts related to avoid useless uses of resources \cite{Auer2011,Arnold2011,EARTH2009, DEDOMENICO2014, MAHAPATRA2013,DeDomenico2012}  when and where the offered system capacity exceeds the momentary and local real need. With 5G, the concept of energy efficiency evolved from a communication network centric to both energy efficiency for user terminal and network operation and, the efficient support of (edge) cloud. With 5G, we experience how the densification of the connect-compute networks brought impressive energy reduction costs \cite{Gandotra2017}. 
Today, while designing 6G, the push for sustainability evolves from pure energy efficiency of communication and support of cloud to four main pillars: (i) energy efficiency through AI assisted communication-computation and control network services and operation, (ii) a goal-oriented energy efficiency of AI mechanism including novel paradigm to compress and exchange data and  knowledge sharing \cite{CalvaneseGOWSC2021} (iii) life cycle extension of materials employed to upgrade or redeploy networks (see deliverable D1.1 of the Hexa-X project \cite{CalvaneseHexaX2021} and, (iv) limiting the impact of electromagnetic fields (EMF) radiation to both intended and not-intended users, for instance thanks to novel paradigms of wireless communication channel  shaping with Reconfigurable Intelligent Surfaces (RIS) \cite{DiRenzoRIS,CalvaneseRISE-6G2021,GeorgeRIS} and EMF aware adaptive mechanism \cite{Lexnet}. This implies the definition of innovative solutions and technologies to reduce the cost associated to the use of resources to accomplish a task or an identified goal. This includes reaching new solutions to achieve \textit{energy free connectivity} though wireless harvesting, zero energy radio, reduction and compression of generated and exchanged data for AI mechanisms. It also requires disruptive solutions on \textit{new technologies} including meta-materials for RIS to enable the new paradigm of \textit{wireless environment as a service} \cite{CalvaneseRISE-6G2021}.

In this paper, we advocate that even if virtualization, softwarizatoin, cloudification and O-RAN approaches introduced with 5G and its evolution allow - in principle - to share hardware and compound functions and services on demand,  there is still a huge gap to fill between what the hardware and technology are expected to provide and, what they can offer in reality. We advocate that today existing hardware and technology are not able to fully support effective evolution toward the next generation of connect-compute-intelligence networks. We point out that today the hardware, technologies and new materials required to effectively meet the unprecedented performance targets  required by future 6G services and network operation, have not been designed, tested or even do not exist yet. 

The contribution of this paper is three-fold. First, this paper presents crucial 6G key performance indicators on hardware and technology design. This is crucial to answer to the question: \textit{how to translate system level 6G vision on KPIs requirements into hardware and technology requirements}? Second, we present a new hardware technologies design methodology conceived to enable the effective software-hardware components integration required to meet the challenging performance envisioned for future 6G networks. Third, we suggest a paradigm shift towards \textit{goal-oriented} and \textit{semantic communications}, in which a totally  new opportunity of joint design of  hardware, artificial intelligence and effective communication is offered. The proposed vision is consolidated by our recent results on hardware, technology and machine learning performance with a detailed example on the benefit of AI to relax constraints on hardware design requirements.

\section{6G KPIs and New Services}
\label{sec:KPI}
Academia, industry and standardization bodies are actively working to shape the vision on \textit{what should be 6G} \cite{network203020196Gusecases}. Candidate Key Performance Indicators (KPIs) have been proposed at the system level for future 6G services and use cases \cite{Calvanese6GVTM2019,CalvaneseGOWSC2021,nakamura20205g,Mourad2020}. A summary of system level KPIs is reported on table \ref{tab:6G-KPIs}.
\vspace{-0.2cm}
\subsection{6G New Services}
At the horizon of 2030, 6G networks will support new types of services. First, services already supported by 5G networks will be enhanced to accommodate for the plethora of new use cases thanks to the already planned evolution of current 5G technologies. This evolutionary paradigm in designing the "next generation" push hardware, software and network architecture design to their extreme performance, targeting to accommodate the exponential growth of generated data and ever increasing multi-fold constraints on services. In addition, as introduced in \cite{CalvaneseGOWSC2021}, 6G will incorporate also the following new services that motivate the need for a new generation of wireless communication networks. Such new 6G services will require disruptive technology to be operational: \\
\textbf{\textit{Massive Machine Type Communications supporting Distributed Intelligence}} (MMTCxDI) services - With this family of services, the massive connectivity to support distributed AI mechanisms is the key service requirement.  Criticality, effectiveness, interoperability and scalability will be main characteristics of this new service class. Examples of future applications include intelligent transportation systems, connected living, smart agriculture, super smart cities, etc. \\
\textbf{\textit{Globally-enhanced Mobile BroadBand}} (GeMBB) services - where extreme Mobile Broadband connectivity is required to provide connect-compute-store and AI support over non-densely populated areas (rural, oceans, sky, etc.) is the key service requirement. This service will expand the computation-oriented communication environment to support applications requiring high capacity connectivity in both dense and remote locations, such as rural areas, oceans, and the sky \cite{Calvanese2020Sky6G}, {\it on demand}, i.e. when and where needed. \\
\textbf{\textit{Ultra-Reliable, Low Latency Computation, Communication and Control}} (URLLCCC) - this class of services will extend the capabilities of Ultra-Reliable, Low Latency Communications (URLLC) services already supported by 5G networks. It incorporates computation services running at the edge of the network and end-to-end (E2E) (remote or automated) control. Here \textit{reliability} does not only address the communication but also on computation, AI support and effectiveness of control mechanism are the key service requirements of this class of services. Therefore, reliability and latency refer not only to the communication aspect, but also to computation and AI support sides. Examples of use cases supported by this new service will include factory automation,  multi-sensory XR \cite{saad2019vision}, connected and autonomous terrestrial and flying vehicles, etc.\\
\textbf{\textit{Semantic Services}} - These services will support all applications involving a share and intertwining of knowledge between the interacting natural and artificial intelligence. These services will require to capture, transmit and process large amount of information between cyberspace and physical space without exceeding the application delay. It will become possible that cyberspace supports humans' thought and action in real-time through wearable devices and micro-devices mounted on the human body. 
Such service will support empathic and haptic communications, affective computing, autonomous bi-directional interactions between different Cyber-Physical-Spaces.  Semantic services will offer \textit{intelligence as a service}, a radical paradigm shift enabling not only to connect things but rather to connect knowledge and reasoning capabilities.

\vspace{-0.2cm}
\subsection{6G KPIs: the System Level and the Technology Perspective}
\begin{figure}[h!]
    \centering
   \includegraphics[width=0.9\columnwidth]{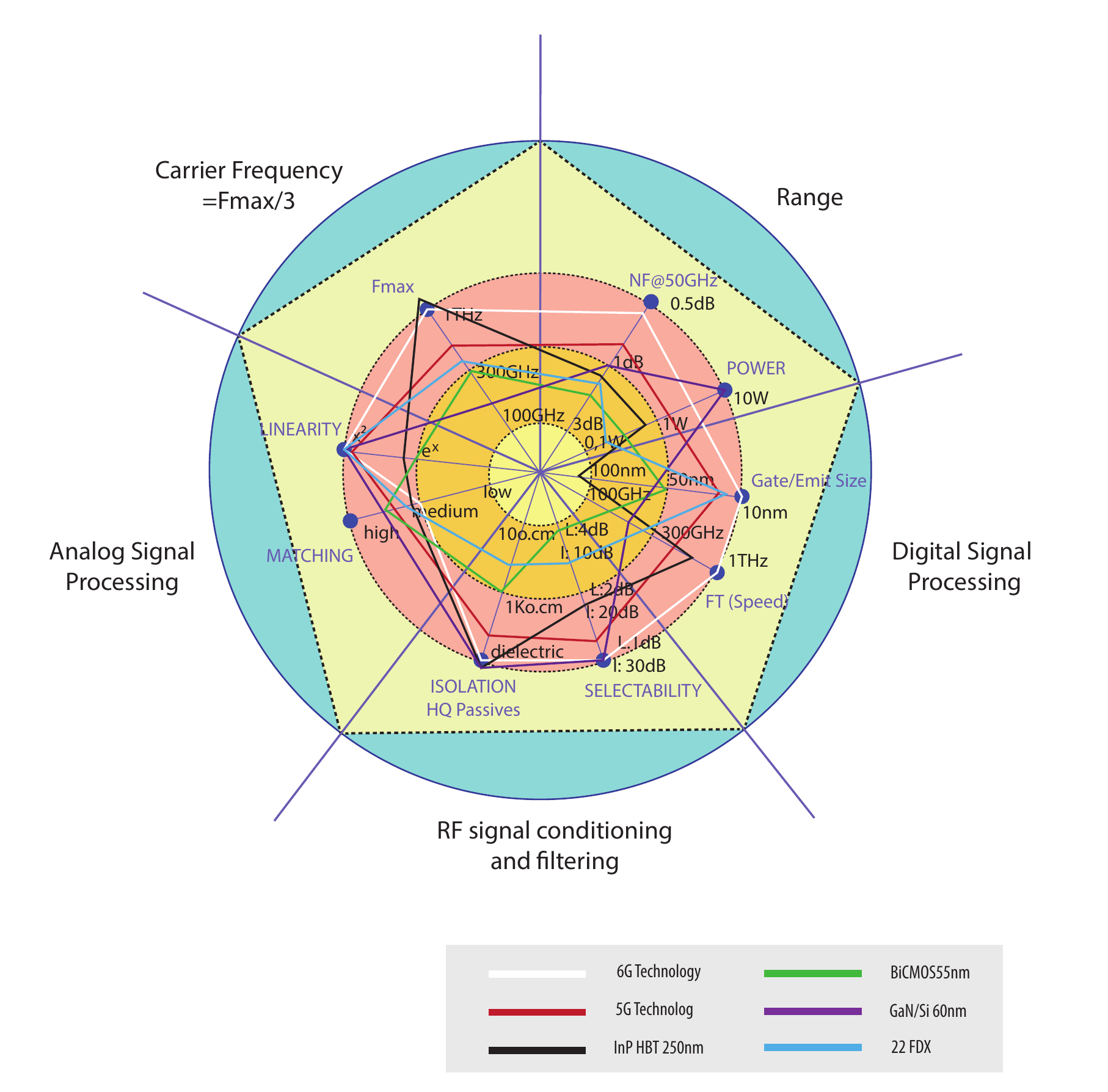}
    \caption{6G hardware KPIs and solid-state technology comparison.}
    \vspace{-0.5cm}
    \label{fig:Techno}
\end{figure}

We propose to consider three classes of 6G KPIs:\\
\textbf{Intrinsically predictive system level KPIs}: The ensure the conceptual continuity with current 5G networks extending performance requirements on already envisaged sets of KPIs. Those KPIs are peak data rate, area traffic capacity, connections density, communication reliability, E2E latency, spectrum efficiency, energy efficiency, etc. (see Table \ref{tab:6G-KPIs}).\\
 \textbf{New system level KPIs}: New use cases, new services and new societal challenges suggest new requirements to the next generation. This entirely new class of KPIs on 6G includes system level performance requirements to support the vision toward a connect-computation-intertwining of intelligence oriented network. Those KPIs are  on  reliability of the decisions taken by intelligent agents supporting network operation and intertwining though the network, the time required to take decisions, the energy efficiency per task or goal to be accomplished, the extension of connect-computer-store-control capabilities to non-terrestrial networks (NTN) integarting terrestrial network services, sustainability of technology, sustainability of network operation,  etc. (see Table \ref{tab:6G-KPIs}).\\
 \textbf{New KPIs on hardware and technology}: KPIs specifying the process technologies and hardware performance required to support KPIs at system level. (see Figure \ref{fig:Techno}).
 
 \vspace{-0.4cm}
\begin{table}[ht!]
\renewcommand{\arraystretch}{1.25}
\setlength{\tabcolsep}{8pt}
\centering
\caption{Comparison of 5G and 6G system level KPIs; NS= Not Specified: TBD= To Be Defined case-by-case. \cite{CalvaneseGOWSC2021}}
\vspace{-0.1cm}
\label{tab:6G-KPIs}
\begin{tabular}{ | l | c | c | } 
\hline
KPI& \textbf{5G} & \textbf{6G} \\ 
\hline \hline
Traffic Capacity & 10 Mbps/$m^2$ &  $\sim$ 1-10 Gbps/$m^3$ \\ 
\hline Data rate DL & 20 Gbps &  1 Tbps \\ 
\hline
Data rate UL & 10 Gbps &  1 Tbps \\ 
\hline
Uniform user experience & \makecell{50 Mbps 2D \\ everywhere} & \makecell{10 Gbps 3D \\ everywhere} \\ 
\hline
Mobility  & 500 Km/h & 1000 Km/h \\
\hline
Latency (radio interface) &  1 msec &  0.1 msec \\ 
\hline
Jitter & NS &  1 $\mu$sec \\
\hline
Communication reliability & $1-10^{-5}$ & $1-10^{-9}$\\
\hline
Inference reliability & NS & TBD\\
\hline
Energy/bit & NS &  1 pJ/bit\\
\hline
Energy/goal & NS &  TBD\\
\hline
Localization precision & 10 cm on 2D &  1 cm on 3D\\
\hline
\end{tabular}
\end{table}
\vspace{-0.2cm}

With this paper we advocate that there is a critical gap to fill between the system level vision on 6G suggesting \textit{system level KPIs} and, what can be effectively done with hardware and technologies available at the horizon of 2030.  A clear vision on future 6G is that applications and services will require up to a factor of one $1000$ in link capacity enhancement. Moreover, we have already witnessed starting from 4G, an incredible increase in the uplink capacity demand \cite{oueis2016uplink} for many services and uses cases involving the support of the cloud, the edge cloud, machine learning and AI mechanisms. As for previous generations, to provide more link capacity it is required to explore new spectrum opportunities. At system level this implies to rethink the network architecture, most probably leading to higher degrees of network densification, higher communication spectral efficiency, more directive communications, new interference management procedures, new adaptive mechanisms (most probably supported by AI), etc. 
The question we address in this section is \textit{how the quest of higher spectrum translate on the need for designing and engineering new hardware and technology processes?}\\ Once thought of as unusable frequencies, the terahertz (THz) bands (above 90 GHz) are one contender for communication technologies applied to 6G, the other being visible light communications (VLC) bands \cite{HaasLifi, haas2020Lifi6G, Calvanese6GVTM2019}.
THz communication system, which modulates base-band signals directly into a continuous THz carrier wave have several scenarios of applications: indoor wireless
mobile networks supporting telepresence though multi-sensory holographic teleportation, video conferencing,  multi-sensory and mobile immersive eXtended reality (XR), extreme capacity Xhaul, nanoscale communication networks for health monitoring via nano-machines, and NTN communications such for inter-satellite and high altitude platforms communication. Although THz communications have obvious advantages for 6G and beyond applications, many technical challenges at both system and hardware technology levels still need to be solved before practical and cost effective deployment of related services.

In order to enable communications above 90 GHz, dedicated high-frequency hardware components have to be designed and solid-state technologies engineered. For solid-state THz communication components and systems, it is difficult to design efficient radio-frequency (RF) circuits such as THz mixers, THz oscillators, THz power amplifiers (PA) and THz antennas. Specifically, it is an open challenge to design ultra-broadband THz antennas (where the communication bandwidth exceed $20 \%$ of the communication carrier frequency) with high (directive) gain and fast beam scanning and reconfiguration functions. As an example, the low noise design for such high bandwidth super-heterodyne transceivers is also a unsolved challenge. Another determining performance factor in the use of THz communication systems which present still unsolved issues is the performance of the THz modulator. Specifically, it is
desirable to design an amplitude THz modulator with high modulation speed and depth, as well as a phase modulator with large scale and linear phase shifts. Today prohibitive complexity and costs of THz communications' hardware might impose a reality check in the real operational adoption of such communication bands. The high hardware cost is due to the extremely large number of RF chains and/or high working frequency band, together with the high energy consumption and severe heat dissipation issue, seriously impede the wide-scale usage of the aforementioned technologies.

\begin{figure}[htb!]
    \centering
   \includegraphics[width=\columnwidth]{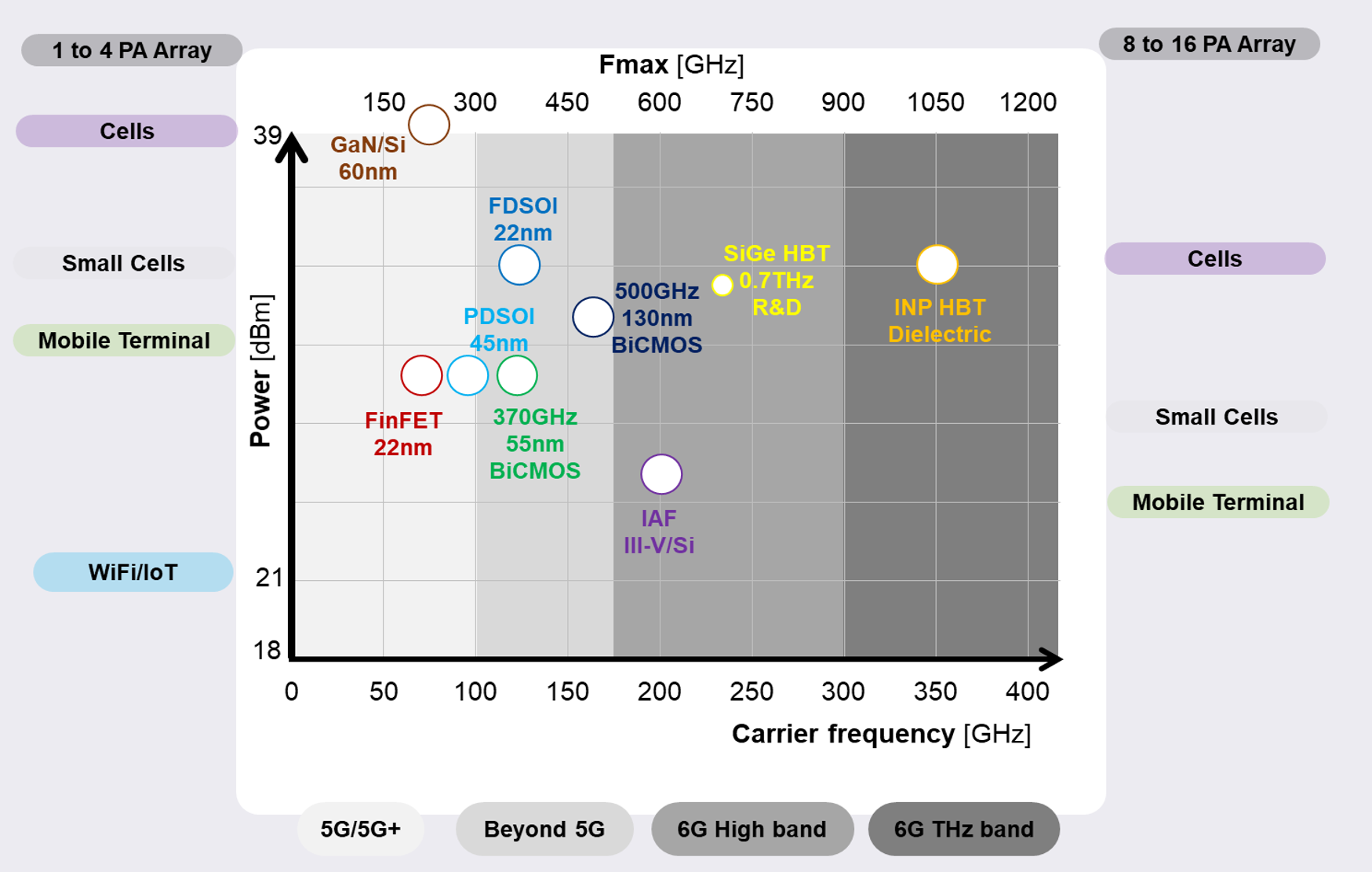}
    \caption{Solid-state process candidates and their limits for communication frequency operating between 60 GHz and 350 GHz.}
    \label{fig:TechnoLimits}
\end{figure}

In Figure \ref{fig:TechnoLimits} we analyse the possible technology candidates for frequency bands between 60GHz up to 350 GHz. At such frequencies, we still ignore which will be the required output power to serve future beyond 5G/6G services. 
Critical to the definition of technology components are:\\
 \textbf{Fmax}: defined as the frequency where the transistor power-gain is 0dB; this parameter impacts the gain availability of receive and transmit chains. Depending on the amplifier architecture, the \textit{Fmax}  impact can be very important. In Figure \ref{fig:TechnoLimits} we see how current technologies impose a practical thumb rule on \textit{Fmax}. In linear class A  power amplifier for the case of PA operating in their linear zone, the maximum carrier frequency ($f_0$) will be lower than \textit{Fmax} /3; When PA exits the linearity zone or for instance in a switch class D amplifier, the carrier frequency should be lower than \textit{Fmax} /10 in an ideal case.  The \textit{Fmax}  parameter defines indeed the maximum carrier frequency of the application.\\
\textbf{Power}: The power availability of a technology depends on break down voltage (BV), and maximum current drove by the transistor (Imax) values, and can be expressed in Watt (W). The covered range for solid-state transistors is between 100mW and 10W. This does not mean that a transmitter cannot aim at more than 10W, but only that the single Power Amplifier using a process technology \textit{Fmax}  transistor cannot provide more than the value presented in the target. \\
\textbf{NF}: The minimal Noise Figure (NF) determines the capacity of the receiver to receive signal with low level of added noise. Power (emitted signal) and NF (received signal) impact directly the wireless link budget, then represent the capacity to cover short or long Range. \\
\textbf{Transistor (gate or Emitter) size}: Digital control of the RF circuit, and digital pre-processing, are intensively used. This means that an “RF process” needs to provide digital solution, which, obviously, is not a “High Performance Computing” one. Two kinds of digital inverters can be sorted out, the classic CMOS, when NFET and PFET transistors are available, and the ECL/CML inverters, more current and size consuming, when only HBTs or NFET are available. \\
\textbf{FT}: Ft gives the potentiality for high frequency digital clock applications and RF oscillators’ application. Efficiency (Transit time / current), is directly issued from FT parameter. The High Speed Digital Integration capability depends on the transistor size and efficiency, these two parameters are defined by the transistor size, the inverter size, and the FT of a process. This will give the potentiality of an efficient and integrated Digital Signal Processing application. \\
\textbf{Selectability}: is the ability to switch RF and mmW signals with high isolation and low loss. This is mainly used to do antenna switch between TX and RX in Time domain duplex mode; or to switch from one mode to another in antenna sharing transceiver. The main specificity of such functionality is the low loss through when the switch is on, and a high isolation when the switch is off. Commonly, this function is done with FET transistors, even if it is feasible with HBT with lower isolation property. \\
\textbf{Isolation and HQ Passives}: are given by substrate resistivity and the presence of thick Metal levels, higher is the application frequency but also higher is the thick metal altitude, lower is the impact of the substrate, this means that this specificity mainly impacts low and medium frequency applications up to 30GHz. Selectability and Isolation HQ Passives impact the RF signal conditioning and filtering application.\\
\textbf{Linearity}: the relation between the output current and the input voltage-control signal of the transistor gives the first order of the linearity. For a FET transistor, this relation is quadratic while for the HBT, it is exponential. In a second and third order, gm2 and gm3 are playing to define IMD2 and IMD3 of the amplifiers. As these gm values are not easy to access, we have focused this analysis at the first order, where a quadratic relation is easier to compensate than an exponential. \\
\textbf{Matching}: this property defines the difference behavior between two minimum size transistors close together in a differential structure. It is better when the transistor size is bigger, or can be compensated as in FDSOI processes using back-gate voltage control. Linearity and Matching are very important in Analog signal processing at Base Band frequency. 

As shown on Figure \ref{fig:TechnoLimits}, \textit{Fmax}  sets the limits on the carrier frequency operation range for each analyzed solid-state process technology. We compare different solid-state technologies in terms of carrier frequency and \textit{Fmax} versus the normalized output power of PA. To provide a fair comparison, we normalize the power as follows. The power for each technology represents the maximum power available with 200mA current through the transistor in a class A cascade PA architecture. 
We consider two sets of transmitter units: PA output power needed for transmitter units  (ex. user mobile terminal, small cells and cells or macro cells) implemented with (1) one to 4 PA on the same device (left side of Figure \ref{fig:TechnoLimits}) and, (2) from 8 to 16 PA  (left side of Figure \ref{fig:TechnoLimits}).

In details, we consider the following solid-state technologies issue from industrial applied research:\\
\textbf{CMOS} technologies, and part of SiGe BiCMOS ones, partially cover the D-Band \textit{Fmax}  requests, 22FDx process being the one, which delivers the higher output power with 28dBm. There is not any CMOS process, which yet satisfies the \textit{Fmax}  request for 2025 Beyond 5G Roadmap \cite{Neired}.\\
\textbf{SiGe BiCMOS} technology processes will cover all the D-Band \textit{Fmax}  requests: for instance, the 500GHz 130nm SiGe HBT from IHP is yet a product, and nowadays, other European companies are developing such 500GHz \textit{Fmax}  SiGe HBT processes. Their power availability is around 27dBm. In addition, their \textit{Fmax}  is close to the needed one, for OH2 absorption peak at 180GHz. This opens the possibility to do Bio imaging and radar applications, with degraded output power, at this frequency, with these SiGe BiCMOS Processes. These  processes satisfy \textit{Fmax}  request for Beyond 5G 2025 Roadmap.
 
Research institutes and academic laboratories are also very active in this vivid domain of research, targeting solutions to design solid-state technologies with \textit{Fmax}  reaching one THz or more. The target of one THz for \textit{Fmax}  is requested for 2030 and beyond 6G connectivity \cite{Neired}. Few examples of process design are given below:\\
Fraunhofer \textbf{IAF III-V / Si} Research process (Figure  \ref{fig:TechnoLimits}, purple line) covers all D-Band and OH2 absorption peak at 180GHz \textit{Fmax}  requests. It, also, partially covers the High part of the G-Band. Its power availability is 24dBm. This process satisfies \textit{Fmax}  request for Beyond 5G 2025 Roadmap, and prepares the next step.\\
\textbf{SiGe BiCMOS 0.7 THz} Research and Development process ( Figure  \ref{fig:TechnoLimits}, yellow line), from IHP, covers all D-Band and OH2 absorption peak at 180GHz \textit{Fmax}  requests. It, also, covers a big part of the High part of the G-Band. Its power availability is 27dBm. This silicon technology is the highest \textit{Fmax}  state of art, and could, (with degraded output power), aim at proposing solutions for the OH2 absorption peak at 325GHz. This process satisfies \textit{Fmax}  request for Beyond 5G 2025 Roadmap, and partially covers 2030 6G connectivity requests. \\
\textbf{InP HBT} Research and Development process (Figure  \ref{fig:TechnoLimits}, orange line), from Teledyne, covers all mmW to THz applications. Its power availability is 28dBm. This process satisfies \textit{Fmax}  request for 2030 6G Conectivity Roadmap.

\begin{figure}[h!]
\vspace{-0.4cm}
\begin{centering}
	\includegraphics[width=\columnwidth]{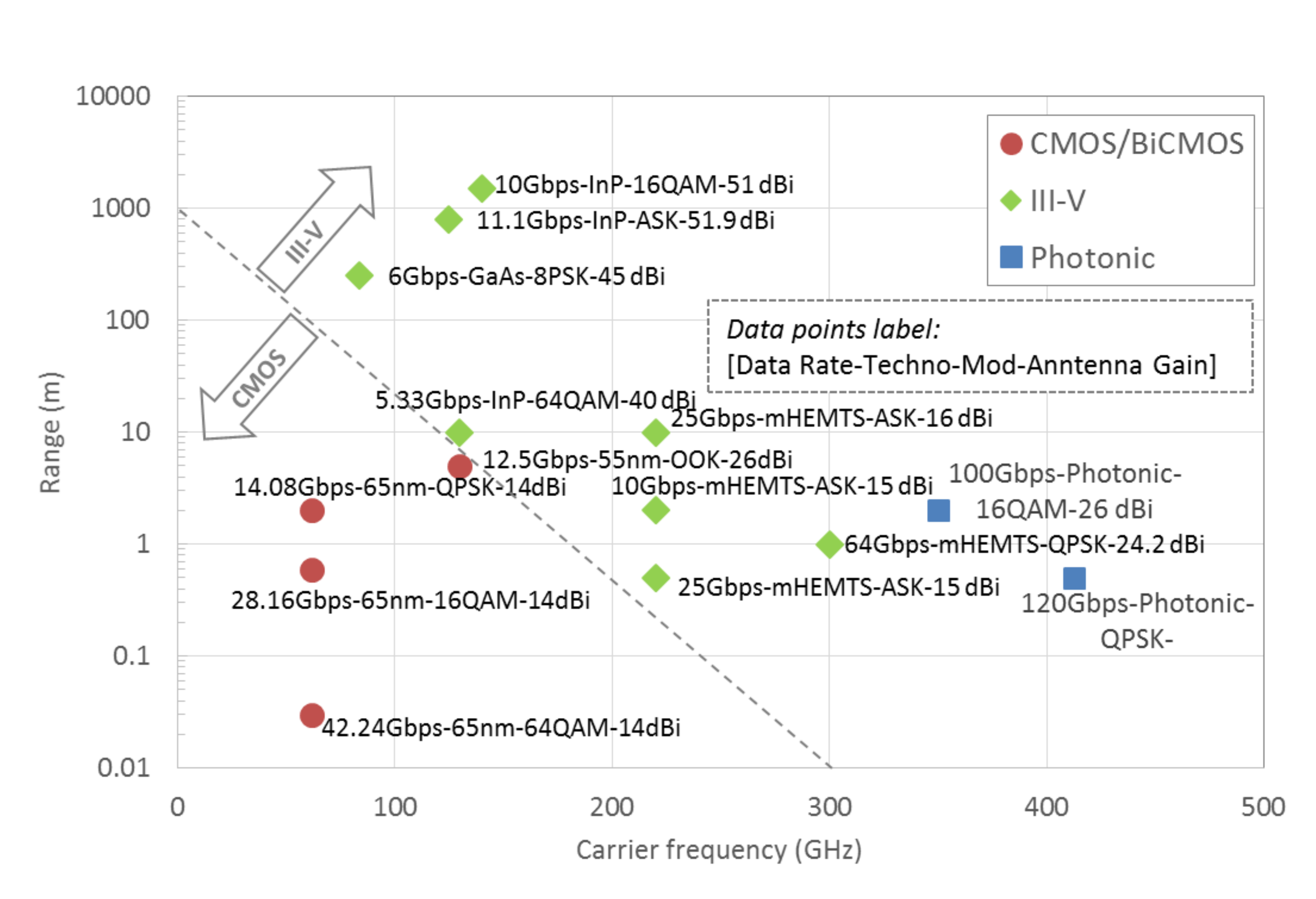}
\caption{Sub-THz Hardware IC Technologies. Courtesy of \cite{Calvanese6GVTM2019}.}
\vspace{-0.2cm}
\label{fig:THz-HW-Tecnologies}
\end{centering}	 
\end{figure}

In summary, each technology can bring its added value in the complete wireless communication physical layer, at different carrier frequency: \\
\textbf{GaN/Si} processes are ongoing to cover FEM requests up to 100GHz in medium term. \\
\textbf{PD-SOI} processes are ongoing to cover FEM request up to 150GHz in medium term. They are well placed for RFFE and Frequency generation circuits up to 150GHz too.\\
\textbf{FDSOI 22nm to 15nm} generations, integrating LDMOS, are and will be very attractive for SOC solutions up to 150 – 200GHz.
In general, \textbf{CMOS} processes will be limited to applications under 200GHz, in medium term. \\
\textbf{SiGe HBT BiCMOS} processes are and will be very attractive for RFFE, FEM, (especially with PIN diodes), Frequency generation and VCO. In medium term, they will cover application up to sub-THz frequency (up to 325GHz). \\
\textbf{GaAs/Si} processes are the main competitors to BiCMOS processes in currently, even if their TRL is still lower. \\
\textbf{InP HBT} processes, are very interesting for long term, and must find a way to be integrated over silicon. \\
In our view a fundamental axe of research will be the design of heterogeneous 3D packaged Transceiver for D-band and beyond connectivity.
In order to further compare the technologies and define the target technology KPIs, we provide a \textit{spider graph} representation of what should be requested to a process technology to full fit with the RF and mmW wireless transceiver challenges, different criteria are presented on Figure \ref{fig:Techno}. 
Figure \ref{fig:Techno} presents with the red line the target performance for full operational 5G networks (at the horizon of 2025) with the black line and, with the white line the target performance for 6G future networks (at the horizon of 2030 and  beyond). In addition in Figure \ref{fig:Techno} we present state of the art representations of the 4 most representative existing technologies (product or R\&D): the 22FDX process, FDSOI 22nm from GF (light blue line); the BiCMOS55nm, SiGe HBT + 55nm CMOS process from ST (green line ); the D006GH, 60nm GaN/Si from OMMIC (purple line) and the R\&D 250nm InP HBT from Teledyne (black line). Each of them has its strenghs and weaknesses, and only the InP HBT allows working at 300GHz, which is the 6G maximum target frequency. The race for concurrent technologies is already declared opened. 
We also show and compare recent implementations of multi-Gbps wireless transceiver using different integrated circuit (IC) technologies. In Figure  \ref{fig:THz-HW-Tecnologies} we report on transceiver technologies proven for carrier frequencies up to 300 GHz and, achieving communication ranges between 1 m up to 1 km (using high directive gain antennas) \cite{Calvanese6GVTM2019}.

Each point on the Figure represents the data-rate, the technology, the modulation scheme and the antenna gain (if available). The Figure confirms the trend we have defined above on the trade-off between carrier frequency and communication range (linked indeed to the PA output power). The larger carrier frequency solutions use a large amount of RF bandwidth in a single RF channel and simple modulations schemes such as Amplitude-Shift Keying (ASK)/ Binary Phase SK (BPSK) which allows for digital-less demodulation. As we can see,  CMOS silicon technology is somewhere limited to 150GHz carrier frequency and less than 10 meters range but enable use higher order modulations such as Quadrature amplitude modulation (QAM), Quadrature PSK (QPSK), 16QAM or even 64QAM. InP technology offers the possibility of working at 300GHz or covering a range bigger than 1Km. CMOS technologies have demonstrated high data-rates at lower carrier frequencies, comparable with those with much larger bandwidth III-V transceiver. In the figure recent photonic-based modulation transceiver performance is also reported. Their performance correspond to the highest carrier frequencies, but their degree of integration if currently very low. We conclude that advanced digital baseband systems data-rate processing capabilities are still in practice limited to few tens of Gbps. Their power consumption is also bonded. In our view, advances on materials and solid-state technology will enable to solve issues related to hardware complexity and its efficiency limitations. For instance recent advances on  materials and device structures such as GaN-HEMT and Graphene are promising for THz modulator design and antennas design \cite{JornetGRAPHENE}
Even if silicon SiGe technology based circuits are not presented in this Figure, we can conclude that also multi technology assembly would be the solution allying THz frequency RF interface High data rate signal processing and high performance computing for AI in the same object. 

\section{AI assisted Hardware and Technology Effectiveness in 6G}

\subsection{Example of Low Complexity Deep Learning for PA Impairment Compensation} 
A major challenge for future 6G hardware design is to achieve cost effective and operational efficient miniaturization of components for transmitters and receivers operating at  sub-Terahertz bands. As discussed in section \ref{sec:KPI},  \textit{Fmax} sets the carrier frequency operation ranges given the selected solid-state process. Higher \textit{Fmax} can be nominally achieved but at the expenses of more complex solid-state processes (see figure \ref{fig:TechnoLimits}). The real operational \textit{Fmax} depends on the actual PA operational performance.  When PA operates in its linear zone, the maximum carrier frequency is upper bounded at \textit{Fmax} /3. Nevertheless, the operational carrier frequency drops to \textit{Fmax} /10 when the PA exits its linearity zone.

Hereafter, we provide a toy example on how low complexity machine learning can contribute to extend the PA linearity zone and indeed, to increase the energy efficiency especially when PA works on  operational high \textit{Fmax} values. 
Let's consider that the communication system integrates a QAM modulation, an OFDM transmitter, a digital predistortion (DPD) based on neural network (NN) techniques and a PA. We characterize the PA with an amplitude distortion function and a phase shift function. We assume a PA derived from a 3GPP Rapp model for communication above 6GHz \cite{rapp3gpp}.
Thereafter, we consider the following input back-off (IBO) definition :
\begin{equation}
    IBO = \frac{P_{1dB,in}}{P_{avg,in}}, \notag
\end{equation}
where $P_{1dB,in}$ corresponds to the input power at the $1dB$ compression point and $P_{avg,in}$ the average input power. Besides, we consider our model to be noise free. 

The goal is to show how the operational linearity zone of PA can be extended adopting a  very low complexity neural network (10 neurons only) with high operation parallelism thanks to a custom architecture. To this end, we design a neural network to specifically tackle amplitude and phase impairments separately through a polar decomposition of the signal. According to \cite{Zayani2014}, tackling AM/AM and AM/PM distortions gives better results which motivates this choice. It results in a dedicated architecture. We design a Conventional Neural Network DPD which is composed of two neural networks respectively using amplitude and phase information of the signal. Each neural network indeed represents a function correcting respectively the amplitude distortion (AM/AM) and the phase distortion (AM/PM). Both neural networks are specifically designed to correct AM/AM and AM/PM distortions respectively. The AM/AM correction works by taking the output amplitude of the PA and predicting the estimated input amplitude. The AM/PM correction works by taking the input amplitude of the PA and predicting the opposite of the phase shift.
Fewer operations are required to find the opposite phase shift because the problem is likely simpler, resulting indeed in lower complexity. The training phase required uses the Indirect Learning Architecure (ILA) \cite{Psaltis1988}. It consists in deriving a post-distorter and placing it before the PA to perform DPD. Learning is performed by optimizing a mean squared error (MSE) loss function using an Adam \cite{Kingma2017adam} optimizer. This training approach is known as ``conventional" since it realizes a simple gradient descent to update the NN parameters.

\begin{figure}
    \centering
    \includegraphics[width=0.93\columnwidth]{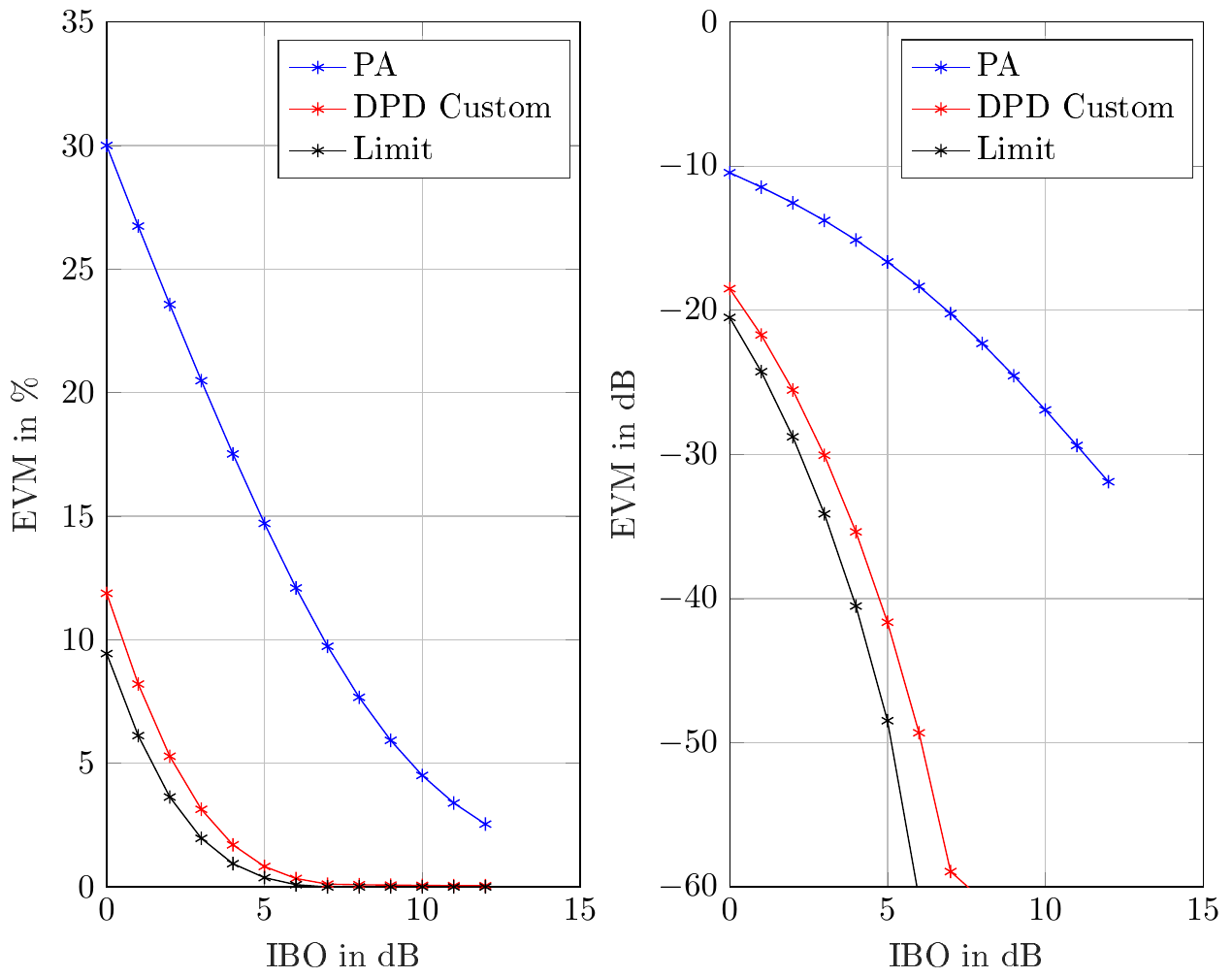}
    \caption{EVM performance using ``conventional" learning}
    \label{fig:evm_ila}
\end{figure}

Our numerical evaluations are shown on Figure \ref{fig:evm_ila}. In the simulation, we assume that parameters of the PA model are fixed according to \cite{rapp3gpp}. Results are represented in terms of Error Vector Magnitude (EVM) in function of the Input Back-Off (IBO). We consider a PA which presents AM/AM and AM/PM distortions. The ``Limit" curve corresponds to a PA linear until its saturation. The performance of our solution is almost optimal using only $10$ neurons which justifies the low-complexity aspect. It must be emphasized that such a solution requires a large amount of data for a single state of our PA, about $10^5$ OFDM symbols which represents a cumbersome database. Indeed, our numerical results show the  benefits of including a neural network  digital predistortion function to compensate possible PA impairments that affect the PA linearity range. 
\vspace{-0,25cm}





\subsection{Support of AI for Efficient Analog-to-Digital Conversion}
\begin{figure}[htb!]
    \centering
    \includegraphics[width=\columnwidth]{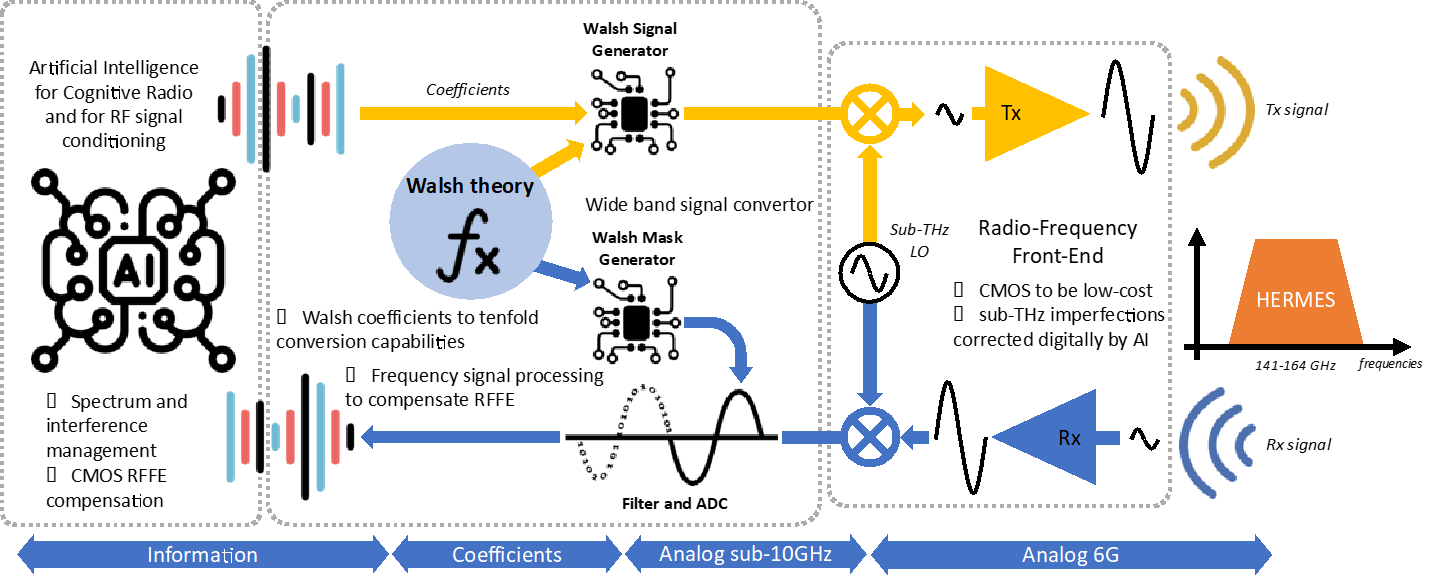}
    \caption{HERMES Project Vision}
    \label{fig:hermesl}
\end{figure}

In order to support ultra-high capacity links at THz frequencies, Analog-to-digital converters (ADC) have to support very high sampling
rate. The hardware energy consumption, chip area and manufacturing costs increase rapidly with their bit resolution. 6G performance is indeed prone to the critical trade-off between ADC resolution and hardware costs.  Consequently, signal processing techniques for low quantized signals must be carefully investigated  for future transceiver architectures.
In this direction, an interesting approach would be to change paradigm in the way the RF chain operates. Currently, the entire signal (modulation, carrier frequency) is processed while the information required to reconstruct the intended signal at destination can be notably compressed. As an example, the H2020 FET Open HERMES project that will kick-off on September 2021,
attempts to rethink the signal generation/reception chain thank to the application of  Walsh theory to wide band conversion from information-to-RF. 
HERMES  goal (see Figure \ref{fig:hermesl}) is to combine for the first time AI, mathematics Walsh theory and CMOS to reach sub-THz frequencies. The solution will be low cost, increasing the energy efficiency by hundreds of percent. It will support actual and future waveforms and give access to large bandwidths instantaneously. 
This allows to select the harmonics of the signal strictly required for correct demodulation of the received signal. In addition, in the proposed methodology, AI agents received as inputs the Walsh series coefficients to enhance the performance of the Radio Frequency Front End (RFFE), adapting the key parameters defining the communication modulation, bandwidth, carrier frequency. This approach can be also advantageously applied to enhance the operational linearity of PA, and the signal-to-noise ratio (SNR) at the receiver. 



\section{Conclusions}

Future 6G will push the network architecture performance to its extreme capabilities. The AI community is expecting future 6G networks to support for a new class of semantic services. This will support applications involving a share and intertwining of knowledge between natural and artificial intelligence. AI is also expected to be applied at all functional levels of the networks to optimize its operation and to reduce its costs. AI will consequently generate and require data to process. Therefore, the data traffic is expected to exponentially grow at both terrestrial and non-terrestrial networks. From these visions, challenging system level KPIs are set. The envisaged direction is to explore new spectrum horizons and target higher spectral efficiency for wireless communications. To this end, new challenging research axes explore the use of sub-Terahertz bands to reach targets of Tbps links capacity. We advocate that there is a critical gap to fill between system level identified ambition and, what the hardware and solid-state technologies will be able to support at the horizon of 2030. Our reality check is that technologies required to meet the challenging system level KPIs of 6G have not been designed, tested or even do not exist yet. Our reality check is that hardware design will be fundamental to meet the rising sustainability and energy efficiency targets. In this paper, we detail our vision on the needed advances required at hardware and technology levels. New technology solutions have to be explored, solving the trade-off between complexity/power consumption of transceiver and antennas and the exploitation of diversity techniques, 
as well as the smart use of AI for both reducing signal processing and data needed for both AI training and communication. A paradigm shift is also expected: the semantic and goal-oriented communication paradigm, where communication and AI support efficiency will be enabled by a new paradigm for exchange compressed data and share knowledge between natural and artificial intelligence agents. To this end, we provide first a comprehensive comparison on solid-state technology candidate. We also discuss their limits. Focusing on the case of sub-Terahertz communications, we identified one of the main issues for transceiver architectures design:  achieving high energy efficiency while enabling the use of \textit{very large communication bands}. This requires to process wide-band signals aggregated from a large number of carriers (more than 10) and potentially, to process in parallel multiple different modulations. Then, we indicate as a possible solutions to leverage on the use of AI to reduce the complexity of the signal processing, to counteract hardware impairments with the support of AI and, to dynamically adapt the setting of hardware to the current electromagnetic communication environment. One conclusion is that multi-technology integration approach will be adopted to meet the challenging requirements on hardware and technology imposed by 6G envisioned services and use cases. Our conclusion is that technology process still requires substantial work to consolidate the opportunity offered by THz communications and that it will be a challenging multidisciplinary field of science.

\bibliographystyle{IEEEtran}
\bibliography{references.bib}

\end{document}